\documentclass[reprint,pra,amsmath,amssymb,aps,showpacs,preprintnumbers]{revtex4-1}
\usepackage{graphicx} 
\usepackage{datetime}
\usepackage{bm} 

\begin{document}
\title{Interacting quantum plasmons in metal-dielectric structures}
\author{Tigran V. Shahbazyan}
\affiliation{Department of Physics, Jackson State University, Jackson, Mississippi 39217 USA}


\begin{abstract} 
We develop a consistent quantum description of surface plasmons  interacting with quantum emitters and external electromagnetic field. Within the framework of macroscopic electrodynamics in dispersive and absorptive medium,  we derive, in the Markov approximation, the canonical Hamiltonian, commutation relations, and coupling parameters for the plasmon modes in metal-dielectric structures of an arbitrary shape whose  characteristic size is well below  the diffraction limit. 
We then develop a new quantum approach bridging the macroscopic and canonical schemes which describes the interacting plasmons in terms of bosonic modes with linear dispersion whose coupling to the electromagnetic field and quantum emitters is mediated by the classical plasmons. By accurately accounting for medium optical dispersion and losses in the interactions of surface plasmons with light and localized electron excitations, this approach can serve as a framework for studying non-Markovian effects in plasmonics. 
\end{abstract}
\maketitle

\section{Introduction}

Over the past decade, quantum plasmonics \cite{maier-np13} underwent a rapid development fueled by a host of recently discovered phenomena such as strong exciton-plasmon coupling effects \cite{bellessa-prl04,sugawara-prl06,wurtz-nl07,fofang-nl08,hakala-prl09,gomez-nl10}, plasmon-assisted hot carrier generation \cite{park-nl11,melosh-nl11,halas-science11}, plasmonic laser (spaser) \cite{bergman-prl03,stockman-natphot08,noginov-nature09,zhang-nature09,gwo-science12}, plasmon tunneling \cite{baumberg-nature12,aizpurua-nl12,dionne-nl13,tan-science14} and more, along with a growing number of applications. Surface plasmons are collective electron excitations living at the metal-dielectric interfaces which can interact strongly with light and localized electron excitations such as excitons in molecules or semiconductors \cite{stockman-review}. Although classical description of many experiments in terms of local field enhancement has  largely been successful, a growing number of topics and applications require a more rigorous quantum approach \cite{hohenester-prb08,nordlander-nl09,stockman-jo10,waks-pra10,garcia-prl13,garcia-optica17,aizpurua-nanophot20}.  

In nanoscale systems, the local fields can change strongly over the length scale well below the diffraction limit, and  so the plasmon interactions with the electromagnetic (EM) field and  excitons depend sensitively on the system  parameters such as the geometry of a metal-dielectric structure and the exciton position relative to it. While  the coupling parameters, characterizing these interactions, have been suggested in several forms by using analogy with the cavity modes \cite{khitrova-nphys06,andreani-prb12,shahbazyan-nl19,mortensen-rpp20}, these parameters have yet to emerge within a consistent quantization approach for interacting  plasmons.  Yet another longstanding  challenge for quantum plasmonics is to  account properly  for strong optical dispersion and losses in metals that give rise to non-Markovian effects \cite{nori-prb09,tejedor-prb10,thanopulos-prb17,moradi-sr18,molmer-acsph19}.

Within the canonical quantization scheme, localized plasmon modes with discrete frequency  spectrum $\omega_{m}$   are described by the Hamiltonian  
\begin{equation}
\label{H-plasmon}
\hat{H}_{\rm pl}=\sum_{m}\hbar\omega_{m}\hat{a}^{\dagger}_{m}\hat{a}_{m},
\end{equation}
where $\hat{a}^{\dagger}_{m}$ and $\hat{a}_{m}$ are, respectively, the plasmon creation and annihilation operators obeying the canonical commutation relations $[\hat{a}_{m},\hat{a}^{\dagger}_{n}]=\delta_{mn}$. Plasmon interactions with quantum emitters (QEs), modeled hereafter by two-level systems, are usually described by the Jaynes-Cummings  interaction Hamiltonian 
\begin{equation}
\label{H-pl-qe}
\hat{H}_{\rm pl-qe}=\sum_{im}\hbar g_{im}(\hat{\sigma}^{\dagger}_{i}\hat{a}_{m}+\hat{a}^{\dagger}_{m}\hat{\sigma}_{i}),
\end{equation}
where  $\hat{\sigma}^{\dagger}_{i} $ and $\hat{\sigma}_{i} $ are, respectively, the raising and lowering operators for the $i$th QE (related to the Pauli matrices for Fermionic systems), while $g_{im}$ is the QE-plasmon coupling, which, within the canonical approach, is an \textit{ad hoc}  parameter. Although widely employed, the canonical scheme has significant limitations when used in metal-dielectric structures characterized by a complex dielectric function $\varepsilon (\omega,\bm{r})=\varepsilon' (\omega,\bm{r})+i\varepsilon'' (\omega,\bm{r})$ since it ignores the medium optical dispersion and, hence, is unsuitable for describing non-Markovian effects in plasmonics.

On the other hand, the medium optical dispersion is accurately accounted for in the macroscopic electrodynamics approach based upon the fluctuation-dissipation (FD) theorem \cite{welsch-pra98,welsch-p00,philbin-njp10}. In this framework, the EM fields are quantized in terms of reservoir noise operators $\hat{\bm{f}}(\omega,\bm{r})$  driven by the Hamiltonian 
%
\begin{equation}
\label{H-noise}
\hat{H}_{N}= \!\int_{0}^{\infty}\! d\omega \!\int\! dV \,\hbar \omega \hat{\bm{f}}^{\dagger}(\omega,\bm{r})\!\cdot\!\hat{\bm{f}}(\omega,\bm{r})
\end{equation}
and obeying the commutation relations 
%
\begin{equation}
[\hat{\bm{f}}(\omega,\bm{r}),\hat{\bm{f}}^{\dagger}(\omega',\bm{r}')]=\bm{I}\delta(\omega-\omega')\delta(\bm{r}-\bm{r}'),
\end{equation}
where $\bm{I}$ is the unit tensor. Interactions with QEs are described by the Hamiltonian term
$\hat{H}_{\rm int}=-\sum_{i}\hat{\bm{p}}_{i} \cdot \hat{\bm{E}}(\bm{r}_{i})$,
where $\hat{\bm{p}}_{i}$ and $\hat{\bm{E}}(\bm{r}_{i})$ are, respectively,  the QE dipole moment and  electric field operators.  The latter is given by
\begin{equation}
\label{field-dissip}
\hat{\bm{E}}(\bm{r})=\!\int_{0}^{\infty}\! d\omega \! \!\int\! dV'\! \bm{D}(\omega;\bm{r},\bm{r}')\hat{\bm{P}}_{N}(\omega,\bm{r}') + \text{H.c.},
\end{equation}
where $\hat{\bm{P}}_{N}(\omega,\bm{r}) = (i/2\pi)\sqrt{\hbar\varepsilon'' (\omega,\bm{r})}\hat{\bm{f}}(\omega,\bm{r})$ is the noise polarization vector operator and  $\bm{D}(\omega;\bm{r},\bm{r}')$ is the EM dyadic Green function defined as 
%
\begin{align}
\label{green-em}
\bm{\nabla}\!\times\! \bm{\nabla}\!\times\!\bm{D}(\omega;\bm{r},\bm{r}')-\frac{\omega^{2}}{c^{2}}\varepsilon(\omega,\bm{r}) \bm{D}(\omega;\bm{r},\bm{r}')~~~~~~
\nonumber
\\
 =\frac{4\pi\omega^{2}}{c^{2}} \bm{I}\delta(\bm{r}-\bm{r}').
\end{align}
The macroscopic FD approach has been extensively used to model spontaneous emission,  strong coupling effects and non-Markovian dynamics in metal-dielectric structures \cite{welsch-pra00,dzsotjan-prb10,hughes-prb12,andreani-prb12,hughes-prb13,zubairy-prb14,garcia-prl14,rousseaux-prb18,sivan-prb19,nori-prb09,tejedor-prb10,thanopulos-prb17,moradi-sr18,molmer-acsph19}.  Its major drawback in relation to plasmonics is that, even though surface plasmons reside primarily at the metal-dielectric interfaces, the eigenstates of $\hat{H}_{N}$ extend over the entire system reservoir, i.e., the Hilbert space,  spanned by the operators $\hat{\bm{f}}(\omega,\bm{r})$, is excessively large. Furthermore, the plasmons only  appear  as resonances in the classical EM Green function $\bm{D}$, so that, in practical terms, this approach is limited to systems that allow  $\bm{D}$ to be evaluated analytically or numerically.

In principle, the  Hamiltonians (\ref{H-plasmon}) and (\ref{H-pl-qe}), along with the canonical commutations relations and QE-plasmon coupling, should be obtained within the macroscopic FD framework starting with a suitable mode expansion for the EM Green function that would define the basis set for normal mode expansion of the electric field operator (\ref{field-dissip}) \cite{varguet-ol16,dzsotjan-pra16,hughes-prl19}. However, for systems of general shape, a straightforward expansion of $\bm{D}$ over a  discrete  set of EM modes that include radiation, such as quasinormal modes \cite{hughes-njp14,hughes-acsphot14,lalanne-pra14,hughes-pra15,muljarov-prb16,lalanne-prb18,lalanne-lpr18}, gives rise to  dissipation-induced coupling between the modes which compicates the   commutations relations \cite{hughes-prl19}. On the other hand, on the length scale well below the diffraction limit, the plasmon interactions with QEs take place primarily via the near field coupling rather than  photon exchange, while the rate of nonradiative losses, determined by $\varepsilon'' (\omega,\bm{r})$, by far exceeds the plasmon radiative decay rate. This implies that in nanoscale systems, the plasmons should be treated as primarily electronic excitations  interacting with the EM field and QEs, and so the quantization of plasmons should be carried on its own, i.e., separate from the radiation field. Furthermore, since the electron motion in metals is unretarded, the plasmon modes in such systems can be described within quasistatic approximation \cite{stockman-review}, which allows, as we show below,  one to define the orthogonal basis set  that is free of  dissipation-induced coupling. 

In this paper, we develop a consistent quantum approach to localized plasmons  interacting with quantum emitters and the electromagnetic field  which accurately accounts for   medium optical dispersion and losses.  First,  starting within the macroscopic FD framework, we employ the near-field plasmon Green function \cite{shahbazyan-prl16,shahbazyan-prb18} do define the normal mode expansion for the plasmon field operators free of dissipation-induced coupling. Using this basis set, we explicitly obtain the plasmon Hamiltonian (\ref{H-plasmon}), along with the equal-time canonical commutations relations, and show that the canonical quantization scheme is valid  only in the Markov approximation, i.e., if the dielectric function $\varepsilon (\omega,\bm{r})$ is replaced by its value at the plasmon frequency (i.e., $\omega=\omega_{m}$). In this way, we also obtain the microscopic coupling parameters that define the plasmon interactions with the EM field and QEs in terms of the plasmon local fields, system geometry and QE positions relative to the plasmonic structure. Second,  moving beyond the Markov approximation, we present a new approach that bridges the macroscopic and canonical quantization schemes while accounting accurately for the medium optical dispersion and losses. In this approach, quantum plasmons are described in terms of a discrete set of bosonic modes with linear dispersion which reside at the metal-dielelectric interfaces, and whose operators span a reduced Hilbert space obtained by projecting the full reservoir states upon localized plasmon modes. The interactions of such projected reservoir modes  with the EM field and QEs are mediated by \textit{classical} plasmons, so that the classical plasmon enhancement effects are encoded in the Hamiltonian coupling parameters. 

\section{From macroscopic to canonical quantization of surface plasmons}

In this section, starting within the macroscopic quantization scheme \cite{welsch-pra98,welsch-p00,philbin-njp10}, we  obtain the canonical Hamiltonian (\ref{H-plasmon}) and the  equal-time commutation relations for localized surface plasmon in metal-dielectric structures of arbitrary shape.  

\subsection{Quasistatic modes and plasmon Green function}
We consider a metal-dielectric structure characterized by dielectric function  of the form $\varepsilon (\omega,\bm{r})=\sum_{i}\theta_{i}(\bm{r})\varepsilon_{i}(\omega)$, where $\theta_{i}(\bm{r})$ is  unit step function that vanishes outside the connected region, metal or dielectric, of volume $V_{i}$ with uniform dielectric function $\varepsilon_{i}(\omega)$.  For unretarded electron motion in metals, the potentials $\Phi_{m}(\bm{r})$ and frequencies  $\omega_{m}$  of plasmon  modes   are determined by the quasistatic Gauss law as \cite{stockman-review} 
%
\begin{equation}
\label{gauss}
\bm{\nabla}\cdot\left [\varepsilon' (\omega_{m},\bm{r})\bm{\nabla} \Phi_{m}(\bm{r})\right ]=0.
\end{equation}
Accordingly, the plasmon mode fields,  which we choose to be real,  are defined as $\bm{E}_{m}(\bm{r})=-\bm{\nabla} \Phi_{m}(\bm{r})$. Importantly, the plasmon mode fields   are orthogonal in \textit{each}  region $V_{i}$, 
%
\begin{equation}
\label{superorthogonality}
\int\! dV_{i} \bm{E}_{m}(\bm{r})\!\cdot\!\bm{E}_{n}(\bm{r})=\delta_{mn}\int\! dV_{i} \bm{E}_{m}^{2}(\bm{r}),
\end{equation}
so that $\int \! dV\varepsilon''(\omega,\bm{r})\bm{E}_{m}(\bm{r})\bm{E}_{n}(\bm{r})=0$ for $m\neq n$, implying no dissipation-induced coupling (see Appendix).

The near-field Green function that defines the  field operator  (\ref{field-dissip}) can be split into free-space and plasmon parts as \cite{shahbazyan-prl16,shahbazyan-prb18} $\bm{D}=\bm{D}_{0}+\bm{D}_{\rm pl}$. When inserted into  Eq.~(\ref{field-dissip}), the first term yields the electric field due to noise fluctuations, while the second term defines  the normal mode expansion of the plasmon field operator. In this paper, we  focus only on the plasmonic sector of the Hilbert space. In the absence of dissipation-induced coupling, the plasmon Green function  can be derived \textit{exactly} in the following form (see Appendix):  
%
\begin{equation}
\label{gauss-green-plasmon-full}
\bm{D}_{\rm pl}(\omega;\bm{r},\bm{r}')=\sum_{m}D_{m}(\omega)\bm{E}_{m}(\bm{r}) \bm{E}_{m}  (\bm{r}')
\end{equation}
where
\begin{align}
\label{gauss-green-plasmon}
D_{m}(\omega) =
\dfrac{4\pi}{\int\! dV \bm{E}_{m}^{2}(\bm{r})} 
-  \dfrac{4\pi}{\int\! dV \varepsilon (\omega,\bm{r})\bm{E}_{m}^{2}(\bm{r})}.
\end{align}
The first  term ensures that $\bm{D}_{\rm pl}=0$ for  $\varepsilon=1$ (or, in the limit $\omega\rightarrow\infty$). The plasmon Green functions exhibits poles in the complex frequency plane defined by the condition $\int\! dV \varepsilon(\omega,\bm{r})\bm{E}_{m}^{2}(\bm{r})=0$. In the following, we restrict ourselves to the plasmonics frequency domain  $\varepsilon''(\omega)/\varepsilon'(\omega)\ll 1$. Since  $\int\! dV \varepsilon'(\omega_{m},\bm{r})\bm{E}_{m}^{2}(\bm{r})=0$ due to the Gauss law, we can expand  $\varepsilon' (\omega,\bm{r})$ in Eq.~(\ref{gauss-green-plasmon}) near $\omega_{m}$ to present the plasmon Green function as a sum over the plasmon poles \cite{shahbazyan-prb18} (see Appendix)
\begin{align}
\label{dyadic-plasmon}
\bm{D}_{\rm pl}(\omega;\bm{r},\bm{r}') 
=
\sum_{m}
\frac{\omega_{m}}{4 U_{m}}\frac{\bm{E}_{m}(\bm{r}) \bm{E}_{m}  (\bm{r}')}{\omega_{m}-\omega -\frac{i}{2}\gamma_{m}(\omega)},
\end{align}
where $U_{m}$ is the plasmon mode energy \cite{landau},
\begin{align}
\label{energy-mode}
U_{m}
= \frac{1}{16\pi} 
\!\int \!  dV     
\dfrac{\partial[\omega_{m}\varepsilon'(\omega_{m},\bm{r})]}{\partial \omega_{m}}
\, \bm{E}_{m}^{2}(\bm{r}),
\end{align}
and $\gamma_{m}(\omega)$ is the frequency-dependent decay rate  
\begin{equation}
\label{mode-decay}
\gamma_{m}(\omega)=\dfrac{2\!\int \!  dV     
\varepsilon''(\omega,\bm{r})\bm{E}_{m}^{2}(\bm{r}) }{\!\int \!  dV     
[\partial\varepsilon'(\omega_{m},\bm{r})/\partial \omega_{m}]
 \bm{E}_{m}^{2}(\bm{r}) }.
\end{equation}
In structures with a single metallic component, 
the decay rate takes the form \cite{stockman-review} $\gamma_{m}(\omega)=2\varepsilon''(\omega)/[\partial\varepsilon'(\omega_{m})/\partial \omega_{m}]$. Finally, with help of Eqs.~(\ref{superorthogonality}) and (\ref{mode-decay}), we obtain the following relation  
\begin{align}
\label{FDT}
\int \! dV  \varepsilon''(\omega,\bm{r}) \bm{D}_{\rm pl}^{*}(\omega;\bm{r},\bm{r}')\bm{D}_{\rm pl}(\omega;\bm{r},\bm{r}'')
~~~~~~~~~~
\nonumber\\
=4\pi \text{Im}\bm{D}_{\rm pl}(\omega;\bm{r}',\bm{r}''),
\end{align}
which ensures consistency with the FD theorem \cite{welsch-pra98,welsch-p00,philbin-njp10}.

\subsection{Plasmon Hamiltonian and canonical commutation relations}

The normal mode expansion of the plasmon field operator is obtained upon inserting the plasmon Green function (\ref{dyadic-plasmon}) into Eq.~(\ref{field-dissip}): 
$\hat{\bm{E}}_{\rm pl}(\bm{r})=\sum_{m}\hat{\bm{E}}_{m}(\bm{r})$, where 
\begin{align}
\label{field-plasmon-normal}
 \hat{\bm{E}}_{m}(\bm{r})
=
\sqrt{\frac{\hbar\omega_{m}}{4 U_{m}}}\,\bm{E}_{m}(\bm{r})   (\hat{a}_{m}+\hat{a}^{\dagger}_{m}),
\end{align}
is  the field operator for an individual mode. Here, $\hat{a}_{m}$ is the plasmon annihilation operator defined as
\begin{equation}
\label{a-plasmon}
\hat{a}_{m}=i\!\int_{0}^{\infty}\! \frac{d\omega}{\sqrt{2\pi}}\frac{\hat{f}_{m}(\omega)}{\omega-\omega_{m} +\frac{i}{2}\gamma_{m}(\omega)},
\end{equation}
where $\hat{f}_{m}$ is  noise operator \textit{projected} on plasmon mode:
\begin{align}
\label{f-proj}
\hat{f}_{m}(\omega)
&=i\sqrt{\frac{\pi\omega_{m}}{2\hbar U_{m}}}\!\int\!dV \bm{E}_{m}(\bm{r}) \! \cdot \! \hat{\bm{P}}_{N}(\omega,\bm{r})
\\
&=-\sqrt{\frac{\omega_{m}}{8\pi U_{m}}}\!\int\! dV\!\sqrt{\varepsilon'' (\omega,\bm{r})}
\,\bm{E}_{m}(\bm{r}) \! \cdot \! \hat{\bm{f}}(\omega,\bm{r}).
\nonumber
\end{align}
Commutation relations for  $\hat{f}_{m}$ follow from those for $\hat{\bm{f}}$ and from Eqs.~(\ref{superorthogonality}) and (\ref{mode-decay}), 
\begin{equation}
\label{f-comm}
[\hat{f}_{m}(\omega),\hat{f}_{n}^{\dagger}(\omega')]=\delta_{mn}\delta(\omega-\omega') \gamma_{m}(\omega),
\end{equation}
where commutativity of operators at $m\neq n$ is insured by the absence of dissipation-induced coupling. Now, using Eqs.~(\ref{a-plasmon}) and (\ref{f-comm}), we obtain the commutation relations for plasmon operators as
\begin{equation}
\label{a-comm}
[\hat{a}_{m}, \hat{a}_{n}^{\dagger}]=\delta_{mn} \!\int_{0}^{\infty}\!  \frac{d\omega }{2\pi}
\frac{ \gamma_{m}(\omega)}{(\omega_{m}-\omega)^{2} +\gamma_{m}^{2}(\omega)/4}.
\end{equation}
In the Markov approximation, by replacing $\gamma_{m}(\omega)$ with $\gamma_{m}\equiv \gamma_{m}(\omega_{m})$  and extending the integral to negative frequencies, we obtain  the  canonical commutation relations  $[\hat{a}_{m}, \hat{a}_{n}^{\dagger}]=\delta_{mn}$. The  plasmon Hamiltonian (\ref{H-plasmon}) follows from  the normal mode expansion (\ref{field-plasmon-normal})   by verifying that the normal-ordered Hamiltonian for individual  modes has the form
\begin{equation}
\label{H-plasmon-mode}
\hat{H}_{m}=\frac{1}{8\pi} 
\!\int \!  dV     
\dfrac{\partial
(\omega_{m}\varepsilon')
}{\partial \omega_{m}}
 \hat{\bm{E}}_{m}^{2}
 =\hbar\omega_{m} \hat{a}_{m}^{\dagger} \hat{a}_{m}, 
\end{equation}
where we dropped the terms $\hat{a}_{m} \hat{a}_{m}$ and $\hat{a}_{m}^{\dagger} \hat{a}^{\dagger}_{m}$ and disregarded the zero-point energy. The factor $1/2$ difference between Eqs.~(\ref{energy-mode}) and (\ref{H-plasmon-mode}) reflects the presence of both positive and negative frequency terms in $\hat{\bm{E}}_{m}(\bm{r})$ \cite{landau}. We stress that, by using the  plasmon Green function (\ref{dyadic-plasmon}), \textit{both}  the Hamiltonian (\ref{H-plasmon})  and canonical commutation relations have been explicitly obtained.

Turning to the plasmon dynamics, the time-evolution of projected noise operators (\ref{f-proj}) is determined by the Heisenberg equations, 
\begin{equation}
\label{noise-dynamics}
\dot{\hat{f}}_{m}(\omega) =-(i/\hbar)[\hat{f}_{m}(\omega), \hat{H}_{N}]=-i\omega \hat{f}_{m}(\omega),
\end{equation}
where the dot stands for   time derivative. From this relation and Eq.~(\ref{a-plasmon}), the Heisenberg  equations  for   plasmon operators  readily follow (in the Markov approximation), 
\begin{equation}
\label{langevin}
\dot{\hat{a}}_{m}(t)  =-(\gamma_{m}/2+i\omega_{m}) \hat{a}_{m}(t) +\hat{f}_{m}(t),
\end{equation}
where 
$\hat{f}_{m}(t)\! = \! (2\pi)^{-1/2}\!\int_{0}^{\infty}\!  d\omega  \hat{f}_{m}(\omega)e^{-i\omega t}$ is time-domain projected noise operator. The commutation relations for $\hat{f}_{m}(t)$ are obtained from Eq.~(\ref{f-comm}) as
\begin{equation}
\label{f-comm-time}
[\hat{f}_{m}(t),\hat{f}_{n}^{\dagger}(t')]= \delta_{mn}\gamma_{m}\delta(t-t'),
\end{equation}
where the Markov approximation was used again. Thus, the Markovian dynamics of plasmon operators $\hat{a}_{m}(t)$ is described by quantum Langevin equation (\ref{langevin}) with  white-noise source $\hat{f}_{m}(t)$,  which  guarantees \cite{scully-book}  the equal-time  commutation relations: $[\hat{a}_{m}(t), \hat{a}_{n}^{\dagger}(t)]=\delta_{mn}$.  
 
\section{Plasmon interactions with  quantum emitters and the electromagnetic field}
 
Consider now the plasmon coupling to QEs and the EM field.  In contrast to cavity modes, the plasmons are  localized at the scale well below the  diffraction limit and, therefore,  interact with the EM field $\bm{\mathcal{E}}(t)$ in a way similar to point dipoles. The interaction Hamiltonian has the form  $H_{\rm pl-em}=-\sum_{m}\hat{\bm{p}}_{m}\cdot \bm{\mathcal{E}}(t)$, where   $\hat{\bm{p}}_{m}=\!\int\! dV\hat{\bm{P}}_{m}(\bm{r})$ is the plasmon dipole moment  and $\hat{\bm{P}}_{m}(\bm{r})$ is the polarization vector operators.  To determine $\hat{\bm{P}}_{m}(\bm{r})$, we present the Gauss law (\ref{gauss}) as $\bm{\nabla}\!\cdot\!\bm{E}_{m}(\bm{r})+4\pi \bm{\nabla}\!\cdot\!\bm{P}_{m}(\bm{r})=0$, where $\bm{P}_{m}(\bm{r})=\chi' (\omega_{m},\bm{r})\bm{E}_{m}(\bm{r})$ is the  mode   polarization vector and $\chi(\omega,\bm{r})$ is the  system  susceptibility. In the Markov approximation, converting this relation to  operator form as $\hat{\bm{P}}_{m}(\bm{r})= \chi' (\omega_{m},\bm{r})\hat{\bm{E}}_{m}(\bm{r})$ and using Eq.~(\ref{field-plasmon-normal}), for the EM field of the form $\bm{\mathcal{E}}(t)=\bm{\mathcal{E}}e^{-i\omega_{L} t}+\bm{\mathcal{E}}^{*}e^{i\omega_{L} t} $ that is uniform on the system scale, we obtain in the rotating wave approximation (RWA)
\begin{equation}
\label{H-pl-em}
H_{\rm pl-em}=-\sum_{m}\left (\bm{\mu}_{m}\cdot\bm{\mathcal{E}}e^{-i\omega_{L} t} \, \hat{a}_{m}^{\dagger} + {\rm H.c.}\right ),
\end{equation}
where  $\bm{\mu}_{m}\equiv \bm{\mu}_{m}(\omega_{m})$, and we introduced the frequency-dependent transition matrix element,
\begin{equation}
\label{mode-matrix-element}
\bm{\mu}_{m}(\omega)=\frac{1}{2}\sqrt{\frac{\hbar\omega_{m}}{U_{m}}}
\!\int\! dV\chi' (\omega,\bm{r})\bm{E}_{m}(\bm{r}).
\end{equation}
to be used in the following section. The scaling factor $\sqrt{\hbar\omega_{m}/U_{m}}$ in Eq.~(\ref{mode-matrix-element})   converts the plasmon  energy $U_{m}$  to $\hbar\omega_{m}$ in order to match the energy of the EM field. With matrix element (\ref{mode-matrix-element}), the plasmon radiative decay rate is given by the standard expression 
%
\begin{equation}
\label{mode-decay-rad}
\gamma_{m}^{rad}= W_{m}^{rad}/U_{m}=(4\omega_{m}^{3} \mu_{m}^{2})/(3\hbar c^{3}),
\end{equation}
where $W_{m}^{rad}=p_{m}^{2}\omega_{m}^{4}/3c^{3}$ is the power radiated by dipole $\bm{p}_{m}$ \cite{scully-book}.

We now turn to  the interactions  between plasmons and QEs   situated at positions $\bm{r}_{i}$  with excitation frequency $\omega_{e}$ and dipole moments $\hat{\bm{p}}_{i}= \bm{\mu}_{i}(\hat{\sigma}^{\dagger}_{i}+\hat{\sigma}_{i})$, where $\bm{\mu}_{i}=\mu_{e} \bm{n}_{i}$   is the transition matrix element ($\bm{n}_{i}$ is the dipole orientation).  Employing the mode expansion (\ref{field-plasmon-normal}) in the interaction Hamiltonian $\hat{H}_{\rm pl-qe}=-\sum_{i}\hat{\bm{p}}_{i}  \cdot \hat{\bm{E}}_{\rm pl}(\bm{r}_{i})$, we obtain, in RWA, the coupling Hamiltonian   (\ref{H-pl-qe})  with   $g_{im}$ given by 
\begin{equation}
\label{coupling}
\hbar g_{im}
=
-\sqrt{\frac{\hbar\omega_{m}}{4U_{m}}}\,\bm{\mu}_{i}\!\cdot\!\bm{E}_{m}(\bm{r}_{i}).
\end{equation}
Using Eq.~(\ref{energy-mode}), the QE-plasmon coupling $g_{im}$ can be recast in the form similar to cavities 
\begin{equation}
\label{coupling-mode-volume}
g^{2}_{im}
=\frac{2\pi \mu_{e}^{2}\omega_{m}}{\hbar{\cal V}^{(i)}_{m}},
~~
\frac{1}{{\cal V}^{(i)}_{m}}
= \frac{2[\bm{n}_{i}\!\cdot\!\bm{E}_{m}(\bm{r}_{i})]^{2}}{\int \! dV [\partial (\omega_{m}\varepsilon')/\partial \omega_{m}]\bm{E}_{m}^{2}},
\end{equation}
where ${\cal V}^{(i)}_{m}$ is the projected  plasmon mode volume  \cite{shahbazyan-acsphot17,shahbazyan-prb18}, which characterizes  the plasmon field confinement at a point $\bm{r}_{i}$ in the direction $\bm{n}_{i}$. Since the Gauss equation (\ref{gauss}) is scale-invariant \cite{stockman-review},  the coupling parameters  (\ref{mode-matrix-element}) and  (\ref{coupling}) are independent of the overall field normalization. By rescaling the fields as $\tilde{\bm{E}}_{m}(\bm{r})=\sqrt{\hbar\omega_{m}/4U_{m}} \bm{E}_{m}(\bm{r})$, these parameters are brought to a more familiar form 
%
\begin{equation}
\label{coupling-normalized}
g_{im}=-\bm{\mu}_{i}\!\cdot\!\tilde{\bm{E}}_{m}(\bm{r}_{i})/\hbar,
~~~
\bm{\mu}_{m}=\!\int\! dV\chi' (\omega_{m},\bm{r})\tilde{\bm{E}}_{m}(\bm{r}).
\end{equation}
%

\begin{figure}[tb]
\begin{center}
\includegraphics[width=0.95\columnwidth]{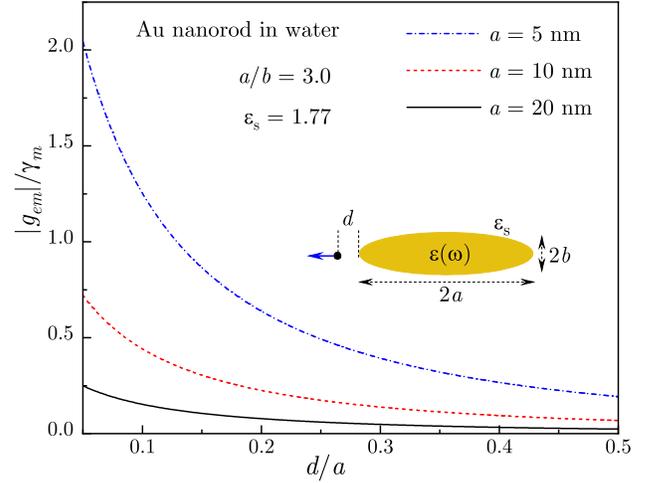}
\caption{\label{fig1} The QE-plasmon coupling  $g_{em}$ for a  QE near the tip of an Au nanorod in water is plotted against the distance to the tip for several values of the nanorod overall size. Inset: Schematics of a QE near  the Au nanorod tip.
 }
\end{center}
\vspace{-3mm}
\end{figure}

To illustrate the sensitivity of QE-plasmon coupling to the system parameters, we  present the results of numerical calculations of the coupling parameter $g_{em}$, given by Eq.~(\ref{coupling-mode-volume}), for a single QE situated at a distance $d$ from the tip of an Au nanorod in water (see Fig.~\ref{fig1}).  The nanorod   was modeled by a prolate spheroid with semi-major and semi-minor axes $a$ and $b$, respectively, at aspect ratio $a/b=3.0$, the standard spherical harmonics were used for calculating the longitudinal plasmon fields, the QE dipole orientation was chosen along the nanorod symmetry axis,  and the Au experimental dielectric function was used in all calculations. The calculated QE-plasmon coupling is normalized by full plasmon decay rate $\gamma_{m}$, which includes the nonradiative decay rate (\ref{mode-decay}) and radiative decay rate (\ref{mode-decay-rad}) taken at plasmon frequency $\omega=\omega_{m}$. The coupling parameter  behaves as $g_{em}\propto 1/\sqrt{{\cal V}_{m}}$, where, for small nanostructures, the plasmon mode volume ${\cal V}_{m}$ scales as the metal volume \cite{shahbazyan-prb18,shahbazyan-nl19}, resulting in the enhanced coupling for smaller nanorods. The coupling sharply increases as  QE approaches hot spot at the nanorod tip characterized by very large plasmon field confinement (small mode volume).

%
%

Finally, the full canonical Hamiltonian for plasmons interacting with the EM field and QEs has the form $H=H_{\rm pl}+H_{\rm pl-qe}+H_{\rm pl-em}+H_{\rm qe}+H_{\rm qe-em}$, where we  included the standard QE Hamiltonian $H_{\rm qe}=\hbar\omega_{e}\!\sum_{i}\hat{\sigma}_{i}^{\dagger}\hat{\sigma}_{i}$  and the coupling term $H_{\rm qe-em}=-\sum_{i}(\bm{\mu}_{i}\!\cdot\!\bm{\mathcal{E}}e^{-i\omega_{L} t} \hat{\sigma}_{i}^{\dagger} + {\rm H.c.})$ describing  the QEs' interactions with the EM field.   We stress that for plasmons, the canonical quantization scheme is valid only in the Markov approximation that ignores the dielectric function dispersion.  However, in metal-dielectric structures, the role of materials optical dispersion can be very significant, and so a quantum description that includes such effects is necessary, as we discuss in the following section.

\section{Beyond Markov approximation: Intermediate quantization scheme}

In this section, we present yet another plasmon quantization approach that bridges the macroscopic and canonical schemes while combining the advantages of both. In this "intermediate" scheme, the Hilbert space is restricted to plasmonic degrees of freedom, in contrast to macroscopic approach, while the medium optical dispersion  is still accurately accounted for,  in contrast to the canonical scheme.

\subsection{Recasting quantum plasmons via projected reservoir modes}

The Langevin equation (\ref{langevin}) with the source (\ref{f-proj}) imply that each plasmon mode is driven  by a fraction of the  reservoir that   overlaps with the plasmon  electric field.  These projected reservoir modes (PRM) form a discrete subspace of the full reservoir Hilbert space  spanned by the operators $\hat{b}_{m}(\omega)=\hat{f}_{m}(\omega)/\sqrt{\gamma_{m}(\omega)}$ or,  using Eqs.~(\ref{mode-decay})  and  (\ref{f-proj}),
\begin{equation}
\label{b-definition}
\hat{b}_{m}(\omega)
=-\frac{\int\! dV\sqrt{\varepsilon'' (\omega,\bm{r})}
\,\bm{E}_{m}(\bm{r}) \! \cdot \! \hat{\bm{f}}(\omega,\bm{r})}
{\left [\int\! dV\varepsilon'' (\omega,\bm{r})
\,\bm{E}_{m}^{2}(\bm{r})\right ]^{1/2}},
\end{equation}
satisfying the commutation relations
\begin{equation}
\label{b-comm}
[\hat{b}_{m}(\omega),\hat{b}_{n}^{\dagger}(\omega')]=\delta_{mn}\delta(\omega-\omega'). 
\end{equation}
Time-evolution in the reduced Hilbert space  is driven by the PRM Hamiltonian  with linear dispersion
\begin{equation}
\label{H-reduced-noise}
\hat{H}_{\rm b}=\sum_{m}  \!\int_{0}^{\infty}\!   d\omega \,\hbar\omega \,\hat{b}^{\dagger}_{m}(\omega)\hat{b}_{m}(\omega), 
 \end{equation} 
which leads to accurate Heisenberg equations  of the form $\dot{\hat{b}}_{m}(\omega)=-(i/\hbar)[\hat{b}_{m}(\omega), \hat{H}_{\rm b}]=-i\omega\hat{b}_{m}(\omega)$ [compare to  Eq.~(\ref{noise-dynamics})]. The PRMs and plasmons can be set as independent variables by adding the  coupling Hamiltonian term:  
$\hat{H}_{\rm pl-b}=i\hbar\sum_{m} \! \int_{0}^{\infty}\! \! d\omega \kappa_{m}(\omega) [\hat{a}^{\dagger}_{m}\hat{b}_{m}(\omega) -\hat{b}_{m}^{\dagger}(\omega)\hat{a}_{m}]$, where  $\kappa_{m}(\omega)=\sqrt{\gamma_{m}(\omega)/2\pi}$ is  the plasmon-PRM coupling. Then, upon tracing the PRMs out, one would arrive, in the standard way, at the master equation for density matrix \cite{scully-book}. Here, we chose a different approach and instead describe the system \textit{directly} in terms of PRMs.

The interaction Hamiltonian between PRMs and QEs is obtained from the QE-plasmon coupling term (\ref{H-pl-qe}) by using the relation  (\ref{a-plasmon}) between  the plasmon and PRM operators, with $\hat{f}_{m}(\omega)=\hat{b}_{m}(\omega)\sqrt{\gamma_{m}(\omega)}$. We then obtain
\begin{equation}
\label{H-b-qe}
\hat{H}_{\rm b-qe}=\sum_{im} \int_{0}^{\infty}\!  d\omega\left [\hbar q_{im}(\omega)\,\hat{\sigma}^{\dagger}_{i}\,\hat{b}_{m}(\omega) +\text{H.c.}
\right ],
\end{equation}
where $q_{im}(\omega)$ is the QE-PRM coupling,
\begin{align}
\label{coupling-prm-qe}
q_{im}(\omega)
= \sqrt{\frac{\gamma_{m}(\omega)}{2\pi}}\frac{ i g_{im} }{\omega -\omega_{m}+\frac{i}{2}\gamma_{m}(\omega)},
\end{align}
with $g_{im}$ given by Eq.~(\ref{coupling}). 

The PRM coupling to the EM field $\bm{\mathcal{E}}(t)$ that is uniform on the system scale is described by the Hamiltonian $H_{\rm int}=-\text{Re} \int dV \hat{\bm{E}}_{\rm pl}(\bm{r})  \cdot  \bm{P}(t,\bm{r})$, where $\bm{P} =\hat{\chi} \bm{\mathcal{E}}$ is the induced polarization vector. For a monochromatic field,  using Eqs.~(\ref{field-plasmon-normal}) and (\ref{a-plasmon}), we obtain
\begin{equation}
\label{H-b-em}
\hat{H}_{\rm b-em}= -\!\!\sum_{m}\! \int_{0}^{\infty}\!\! \! d\omega \! \left [\bm{d}_{m}^{*}(\omega_{L},\omega)\!\cdot \!\bm{\mathcal{E}}e^{-i\omega_{L}t}\, \hat{b}_{m}^{\dagger}(\omega) +\text{H.c.}
\right ],
\end{equation}
where $\bm{d}_{m}(\omega_{L},\omega)$ is the optical transition matrix element for PRMs [compare to Eq. (\ref{coupling-prm-qe})],
\begin{align}
\label{coupling-prm-em}
\bm{d}_{m}(\omega_{L},\omega)
= \sqrt{\frac{\gamma_{m}(\omega)}{2\pi}}\frac{i\bm{\mu}_{m}(\omega_{L})}{\omega-\omega_{m}+\frac{i}{2}\gamma_{m}(\omega)},
\end{align}
and the plasmon transition matrix element $\bm{\mu}_{m}(\omega)$ is given by Eq. (\ref{mode-matrix-element}). 

\subsection{Classical plasmonic enhancement effects and first-order transition probability rates}

To elucidate the mechanism behind the QE-PRM interaction, let us evaluate the first-order transition probability rate for PRM excitation by a QE. For an excited QE with energy $\hbar \omega$, the transition rate for an individual PRM has the form
\begin{equation}
\label{rate-abs}
\Gamma_{im}(\omega)=\frac{2\pi}{\hbar}
\int_{0}^{\infty}\!  d\omega' 
\left | \hbar q_{im}(\omega')\right |^{2}\delta(\hbar\omega'-\hbar\omega),
\end{equation}
where the  integration is taken over the PRM's final states. Evaluating the  integral, we obtain
\begin{equation}
\label{rate-et-mode}
\Gamma_{im} (\omega) 
= 2\pi \left | q_{im}(\omega) \right |^{2}=\frac{g_{im}^{2} \gamma_{m}(\omega)}{(\omega_{m}-\omega)^{2} +\frac{1}{4}\gamma_{m}^{2}(\omega)},
\end{equation}
where we used Eq.~(\ref{coupling-prm-qe}). The full transition rate is obtained by summing Eq.~(\ref{rate-et-mode}) over all PRMs: $\Gamma_{i} (\omega)=\sum_{m}\Gamma_{im} (\omega)$. In fact, the above expression represents the rate of energy transfer (ET) from a QE to plasmons evaluated, in a standard way,   using the classical plasmon Green function (\ref{dyadic-plasmon}) with help of Eq.~(\ref{coupling})  \cite{shahbazyan-prl16},
\begin{equation}
\label{rate-et}
\Gamma_{i} (\omega)
=\dfrac{2}{\hbar}\, \text{Im} \left [\bm{\mu}_{i} \bm{D}_{\rm pl}(\omega;\bm{r}_{i},\bm{r}_{i})\bm{\mu}_{i} \right ] =\sum_{m}\Gamma_{im} (\omega),
\end{equation}
which indicates that QE-PRM interactions are \textit{mediated by classical plasmons} absorbing the QE energy. Thus, the classical effect of resonance ET from a QE to plasmons emerges from the Hamiltonian (\ref{H-b-qe}) in the lowest order.

Turning to PRM interactions with the EM field,  the transition probability rate for excitation of a PRM  by the incident monochromatic  light of frequency $\omega$ is
\begin{equation}
\label{rate-abs-mode}
\Gamma_{m}(\omega)=\frac{2\pi}{\hbar}
\!\int_{0}^{\infty}\!  d\omega' 
\left |  \bm{d}_{m}(\omega,\omega')\!\cdot \!\bm{\mathcal{E}}\right |^{2}\delta(\hbar\omega'-\hbar\omega),
\end{equation}
which, after evaluating the integral, can be presented as [compare to Eqs.~(\ref{rate-et}) and (\ref{rate-et-mode})]
\begin{equation}
\label{rate-abs-pol}
\Gamma_{m} (\omega)
=\dfrac{2\pi}{\hbar^{2}}\left |  \bm{d}_{m}(\omega,\omega)\!\cdot \!\bm{\mathcal{E}}\right |^{2}
=\dfrac{2}{\hbar}\, \text{Im} \left [\bm{\mathcal{E}}^{*}\bm{\alpha}_{m}(\omega)\bm{\mathcal{E}} \right ].
\end{equation}
%
Here, $\bm{\alpha}_{m} (\omega)$ is the optical polarizability tensor of a plasmon mode that defines its response to an external field \cite{shahbazyan-prb18} (see Appendix):
\begin{equation}
\label{polar-mode}
\bm{\alpha}_{m} (\omega)=\frac{1}{\hbar}\frac{\bm{\mu}_{m}(\omega)\bm{\mu}_{m}(\omega)}{\omega_{m}-\omega -\frac{i}{2}\gamma_{m}(\omega)}.
\end{equation}
The full absorption rate is obtained by summing over all PRM modes  as $\Gamma_{\rm pl} (\omega)
=(2/\hbar)\text{Im} \left [\bm{\mathcal{E}}^{*}\bm{\alpha}_{\rm pl}(\omega)\bm{\mathcal{E}} \right ]$, where $\bm{\alpha}_{\rm pl}(\omega)=\sum_{m}\bm{\alpha}_{m}(\omega)$ is the full optical polarizability of plasmonic structure. Within RWA, the absorbed power is given by $P(\omega)=\hbar\omega \Gamma_{\rm pl}(\omega)/4=(\omega/2)\text{Im} \left [\bm{\mathcal{E}}^{*}\bm{\alpha}_{\rm pl}(\omega)\bm{\mathcal{E}} \right ]$. Upon normalizing it by the incident flux $S_{0}=(c/8\pi) \mathcal{E}^{2}$, we obtain the absorption cross-section as $\sigma_{\rm pl}^{\rm abs}=\sum_{m}\sigma_{m}^{\rm abs}$, where $\sigma_{m}^{\rm abs}(\omega)$ is the absorption cross-section for an individual mode,
\begin{equation}
\label{cross-abs-mode}
\sigma_{m}^{\rm abs} 
=\dfrac{4 \pi \omega}{c} \text{Im} \left [\bm{\varepsilon}\bm{\alpha}_{m}(\omega)\bm{\varepsilon} \right ]
=\dfrac{2 \pi \omega}{\hbar c}\frac{\bm{|\varepsilon}\!\cdot\!\bm{\mu}_{m}(\omega)|^{2}\gamma_{m}(\omega)}{(\omega-\omega_{m})^{2}+\frac{1}{4}\gamma_{m}^{2}(\omega)},
\end{equation}
and $\bm{\varepsilon}$ is the incident light polarization.Thus, the PRM-EM interaction is mediated by  plasmon  absorption of the incident light, while the classical absorption cross-section is obtained from the interaction Hamiltonian (\ref{coupling-prm-em}) in the lowest order. Obviously, if the incident light frequency $\omega$ is close to a plasmon mode frequency $\omega_{m}$, this interaction is enhanced due to resonant plasmon absorption. 

\begin{figure}[tb]
\begin{center}
\includegraphics[width=0.95\columnwidth]{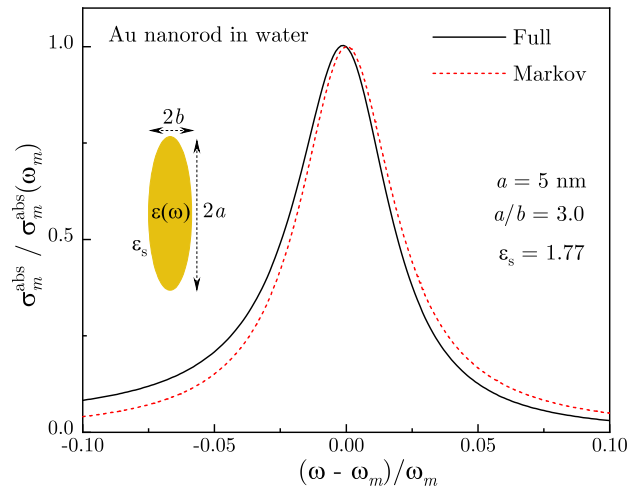}
\caption{\label{fig2} The calculated absorption cross-section of an Au nanorod in water, given by the full expression (\ref{cross-abs-mode}), is compared to the Markov approximation result obtained from the same expression but with  $\omega=\omega_{m}$ in the Au dielectric function. Inset: Schematics of the Au nanorod.
 }
\end{center}
\vspace{-6mm}
\end{figure}

To illustrate the role of optical dispersion in the above first-order effects, in Fig.~\ref{fig2} we compare the normalized absorption cross-section for an Au nanorod in water, calculated using Eq.~(\ref{cross-abs-mode}), with the result of Markov approximation obtained by setting $\omega=\omega_{m}$ in the coupling parameters  $\bm{\mu}_{m}(\omega)$ and $\gamma_{m}(\omega)$. With optical dispersion included, the plasmon resonance spectral shape deviates slightly from the Lorentzian  by exhibiting a red shift of the peak position along with overall enhancement of the lower frequency range. Although, in this case, the effect appear to be relatively small, much stronger non-Markovian effects are expected in higher orders, including in strongly-coupled QE-plasmon systems to be discussed elsewhere.

%
%

The Hamiltonian $\hat{H}=\hat{H}_{\rm b}+\hat{H}_{\rm b-qe}+\hat{H}_{\rm b-em}+\hat{H}_{\rm qe}+\hat{H}_{\rm qe-em}$ provides a starting point for studying quantum correlations and non-Markovian dynamics in hybrid plasmonic systems involving localized plasmons interacting with QEs and the EM field. Within this framework, classical plasmons are encoded  in the coupling parameters (\ref{coupling-prm-em}) and (\ref{coupling-prm-qe}), respectively,  to mediate resonant coupling between the system components.  Namely, the classical  effects of  resonant plasmon excitation by  the EM field and resonance ET between the QEs and plasmons, which underpin most of the plasmon-enhanced spectroscopy phenomena,  now emerge in the lowest order of quantum perturbation theory. In higher orders, these classical effects will modulate quantum correlations and non-Markovian dynamics in hybrid plasmonic systems.

\section{Conclusions}

In summary, we developed a quantization approach for surface plasmons in metal-dielectric structures that  accounts for optical dispersion of the medium complex dielectric function and, hence, is suitable for describing non-Markovian effects in quantum plasmonics. Starting within the macroscopic quantization framework based on the dissipation-fluctuation theorem, we derived, in the Markov approximation, the canonical quantization scheme for interacting surface plasmons  and obtained the relevant coupling parameters in terms of local fields and system geometry.  Beyond Markov approximation, we developed a quantum approach in terms of discrete set of bosonic modes with linear dispersion whose interactions with  the electromagnetic field and quantum emitters are mediated by the classical plasmons, providing resonant coupling between the system components. We have shown that, within this approach, the classical plasmonic enhancement effects, such as resonant plasmon excitation by incident light and resonance energy transfer from a quantum emitter to  plasmonic structure, are obtained in the first-order of perturbation theory.  The quantum dynamics of bosonic modes is restricted to the reduced Hilbert space of reservoir states which overlaps with the electric fields of plasmon modes.

\acknowledgments
This work was supported in part by the National Science Foundation  Grants  Nos. DMR-2000170, DMR-1856515,  DMR-1826886 and HRD-1547754.

 \appendix

 \section{Plasmon modes}

We consider a metal-dielectric structure supporting surface plasmons that are localized at the length scale much smaller than the radiation wavelength. In the absence of retardation effects, each connected volume $V_{i}$ of the structure, metallic or dielectric,  is characterized by a uniform dielectric function $\varepsilon_{i}(\omega)$ so that the full dielectric function  has the form $\varepsilon (\omega,\bm{r})=\sum_{i}\theta_{i}(\bm{r})\varepsilon_{i}(\omega)$, where $\theta_{i}(\bm{r})$ is  unit step function that vanishes outside $V_{i}$. The system eigenmodes are determined by the quasistatic Gauss law \cite{stockman-review}, 
%
\begin{equation}
\label{gauss-law}
\bm{\nabla}\!\cdot\!\left [\varepsilon' (\omega_{m},\bm{r})\bm{\nabla} \Phi_{m}(\bm{r})\right ]=0,
\end{equation}
where $\Phi_{m}(\bm{r})$ and $\omega_{m}$  are the mode potentials and frequencies, respectively, and the mode electric fields, which can be chosen real, are defined as $\bm{E}_{m}(\bm{r})=-\bm{\nabla} \Phi_{m}(\bm{r})$. In the plasmon frequency region, where $\varepsilon''(\omega)/\varepsilon'(\omega)\ll 1$, the mode frequencies are defined by the real part of dielectric function, while its imaginary part defines the mode decay rates.

Let us show that the eigenmodes of Eq. (\ref{gauss-law}) are orthogonal in \textit{each} connected volume $V_{i}$: 
\begin{equation}
\label{orthogonality}
\int\! dV_{i} \bm{E}_{m}(\bm{r})\!\cdot\!\bm{E}_{n}(\bm{r})=\delta_{mn}\int\! dV_{i} \bm{E}_{m}^{2}(\bm{r}).
\end{equation}
Using  $\varepsilon (\omega,\bm{r})=1+4\pi \chi (\omega,\bm{r})=1+4\pi \sum_{i}\chi_{i} (\omega)\theta_{i}(\bm{r})$, where $\chi$ is the susceptibility, we multiply Eq.~(\ref{gauss-law}) by $\Phi_{n}(\bm{r})$ and integrate  over the system volume to obtain
\begin{equation}
\label{orthogonality1}
\int\! dV \bm{E}_{m}\!\cdot\!\bm{E}_{n}+4\pi \sum_{i}\chi'_{i} (\omega_{m})\! \int\! dV_{i} \bm{E}_{m}\!\cdot\!\bm{E}_{n}=0
\end{equation}
Making a replacement $m\leftrightarrow n$  in Eq.~(\ref{orthogonality1}) and subtracting the result from Eq.~(\ref{orthogonality1}), we arrive at  overcomplete system
\begin{equation}
\label{orthogonality2}
\sum_{i}[\chi'_{i} (\omega_{m})-\chi'_{i} (\omega_{n})]\! \int\! dV_{i} \bm{E}_{m}\!\cdot\!\bm{E}_{n}=0,
\end{equation}
and the orthogonality relation Eq.~(\ref{orthogonality})  readily follows. An important consequence of Eq.~(\ref{orthogonality}) is the absence of dissipation-induced coupling between the modes, i.e., for $m\neq n$,
\begin{equation}
\label{dissipation-coupling}
\int \! dV\varepsilon''(\omega,\bm{r})\bm{E}_{m}(\bm{r})\bm{E}_{n}(\bm{r})=\sum_{i} \varepsilon_{i}''\int \! dV_{i}\bm{E}_{m}\bm{E}_{n}=0,
\end{equation}
which allows one to obtain the \textit{exact} plasmon Green function in the presence of losses.

\section{Plasmon Green function}

The EM dyadic Green function for Maxwell equations in the presence of inhomogeneous medium satisfies 
\begin{equation}
\left [\bm{\nabla}\!\times\! \bm{\nabla}\!\times -\frac{\omega^{2}}{c^{2}}\varepsilon(\omega,\bm{r}) \right ] \bm{D} (\omega;\bm{r},\bm{r}') =\frac{4\pi\omega^{2}}{c^{2}}\bm{I}\delta(\bm{r}-\bm{r}')
\end{equation}
where we adopted normalization convenient in the near field limit. Applying $\bm{\nabla}$ to both sides, one finds equation for the longitudinal part of the Green function 
\begin{equation}
\bm{\nabla}[\varepsilon(\omega,\bm{r}) \bm{D} (\omega;\bm{r},\bm{r}')]= -4\pi \bm{\nabla}\bm{I}\delta(\bm{r}-\bm{r}').
\end{equation}
In the near field, it is convenient to switch to the Green function for the potentials $D(\omega;\bm{r},\bm{r}')$, defined as $\bm{D} (\omega;\bm{r},\bm{r}')=\bm{\nabla}\bm{\nabla}'D(\omega;\bm{r},\bm{r}')$, which satisfies 
\begin{equation}
\label{gauss-green-pot}
\bm{\nabla}\!\cdot\!\left [\varepsilon (\omega,\bm{r})\bm{\nabla}D(\omega;\bm{r},\bm{r}')\right ]=4\pi \delta(\bm{r}-\bm{r}').
\end{equation}
%
%
In  free space ($\varepsilon=1$), the near-field Green's function has the form $D_{0}(\bm{r}-\bm{r}')=-1/|\bm{r}-\bm{r}'|$. For arbitrary  $\varepsilon (\omega,\bm{r})$, we separate out the free-space and plasmon parts as  $D=D_{0}+D_{\rm pl}$ to obtain the equation for $D_{\rm pl}$:
\begin{align}
\label{gauss-green-plas}
\bm{\nabla}\!\cdot\!\bigl[\varepsilon (\omega,\bm{r})\bm{\nabla}
&
D_{\rm pl}(\omega;\bm{r},\bm{r}')\bigr]
\nonumber\\
&
=
-\bm{\nabla}\!\cdot\!\bigl [[\varepsilon (\omega,\bm{r})-1]\bm{\nabla}D_{0}(\omega;\bm{r},\bm{r}')\bigr ].
\end{align}
Assume, for a moment, that  the dielectric function $\varepsilon (\omega,\bm{r})$ is real ($\varepsilon''=0$) and expand the plasmon Green's function in terms of eigenmodes of Eq.~(\ref{gauss-law}) as 
\begin{equation}
\label{green-exp}
D_{\rm pl}(\omega;\bm{r},\bm{r}')=\sum_{m}D_{m}(\omega)\Phi_{m}(\bm{r})\Phi_{m}(\bm{r}'),
\end{equation}
with real coefficients $D_{m}(\omega)$. Let us apply to both sides of Eq.~(\ref{gauss-green-plas})  the integral operator $\int\! dV'\Phi_{m}(\bm{r}')\Delta'$. Using the mode orthogonality,  it is easy to prove the relation
\begin{equation}
\int\! dV'\Phi_{m}(\bm{r}')\Delta'D_{\rm pl}(\omega;\bm{r},\bm{r}')= - D_{m} \Phi_{m}(\bm{r}) \!\int\! dV \bm{E}_{m}^{2}(\bm{r}) 
\end{equation}
to use in the left-hand side, and the relation
\begin{equation}
\int\! dV'\Phi_{m}(\bm{r}')\Delta'D_{0}(\omega;\bm{r},\bm{r}')= 4\pi\Phi_{m}(\bm{r})
\end{equation}
to use in the right-hand side.  Then, we obtain
\begin{equation}
\label{gauss-green-plas2}
D_{m} \bm{\nabla}\!\cdot\!\bigl[\varepsilon (\omega,\bm{r})\bm{\nabla}
\Phi_{m}(\bm{r})\bigr]=  
4\pi \dfrac{\bm{\nabla}\!\cdot\!\bigl [[\varepsilon (\omega,\bm{r})-1]\bm{\nabla}\Phi_{m}(\bm{r})\bigr ]}{\int\! dV \bm{E}_{m}^{2}(\bm{r})}.
\end{equation}
Finally, multiplying Eq.~(\ref{gauss-green-plas2}) by $\Phi_{m}(\bm{r})$ and integrating  the result over the system volume, we obtain \cite{shahbazyan-prb18}
\begin{equation}
\label{mode-coeff}
D_{m}(\omega)= \dfrac{4\pi}{\int\! dV \bm{E}_{m}^{2}(\bm{r})} -  \dfrac{4\pi}{\int\! dV \varepsilon (\omega,\bm{r})\bm{E}_{m}^{2}(\bm{r})},
\end{equation}
%
and the plasmon  Green function  takes the form
%
\begin{equation}
\label{gauss-green-field}
\bm{D} (\omega;\bm{r},\bm{r}')=\sum_{m}D_{m}(\omega)\bm{E}_{m}(\bm{r})\bm{E}_{m}(\bm{r}').
\end{equation}
The first  term in Eq.~(\ref{mode-coeff}) ensures that $D_{m}=0$ in the limit $\omega\rightarrow\infty$ (or, in free space with $\varepsilon=1$).

 To incorporate the losses, we note that  in Eq.~(\ref{gauss-green-plas}) with complex dielectric function $\varepsilon(\omega,\bm{r})=\varepsilon'(\omega,\bm{r})+i\varepsilon''(\omega,\bm{r})$, the imaginary part can be considered as a perturbation. In the first order, according to the standard perturbation theory, the diagonal matrix element $\int \! dV\varepsilon''(\omega,\bm{r})\bm{E}_{m}^{2}(\bm{r})$ affects the spectrum but leaves the eigenmodes unchanged, which is equivalent to having full complex dielectric function $\varepsilon(\omega,\bm{r})$ in Eq.~(\ref{mode-coeff}). In higher orders, both the spectrum and the eigenmodes should change as the perturbation causes transitions between the basis states via non-diagonal terms $\int \! dV\varepsilon''(\omega,\bm{r})\bm{E}_{m}(\bm{r})\bm{E}_{n}(\bm{r})$ with $m\neq n$. However, for quasistatic modes, all non-diagonal matrix elements \textit{vanish} [see Eq.~(\ref{dissipation-coupling})], implying that the plasmon Green function Eq.~(\ref{gauss-green-field}) with \textit{complex} coefficients (\ref{mode-coeff}) is \textit{exact} in all orders.

\section{Plasmon pole expansion}

For real $\varepsilon(\omega,\bm{r})$, due to the Gauss law (\ref{gauss-law}), the Green function (\ref{gauss-green-field}) with coefficients (\ref{mode-coeff})  develops a pole as $|\omega|$ approaches $\omega_{m}$. For a complex dielectric function, the plasmon poles move  to the lower half of the complex-frequency plane, and so the Green's function, being analytic in the entire complex-frequency plane except those poles, can be presented as a sum over all plasmon poles. 
For $\omega$ approaching $\omega_{m}$, we expand $\varepsilon'(\omega,\bm{r})$ near $\omega_{m}$ 
\begin{equation}
\varepsilon'(\omega,\bm{r})\approx \varepsilon' (\omega_{m},\bm{r})+ \dfrac{ \partial \varepsilon' (\omega_{m},\bm{r})}{\partial \omega_{m}^{2}}\left (\omega^{2}-\omega_{m}^{2}\right ),
\end{equation}
where we used $\varepsilon'(\omega,\bm{r})=\varepsilon'(-\omega,\bm{r})$, and so the coefficient (\ref{mode-coeff}), after omitting the non-resonant term, becomes
\begin{equation}
\label{mode-coeff2}
D_{m}(\omega) =\frac{\omega_{m}}{4U_{m}}\dfrac{2\omega_{m}}{\omega_{m}^{2}-\omega^{2} -i\omega_{m}\gamma_{m}(\omega)}.
\end{equation}
Here, we introduced the plasmon mode energy \cite{landau}
\begin{align}
\label{energy-mode-plas}
U_{m}
= \frac{1}{16\pi} 
\!\int \!  dV     
\dfrac{\partial[\omega_{m}\varepsilon'(\omega_{m},\bm{r})]}{\partial \omega_{m}}
\, \bm{E}_{m}^{2}(\bm{r}),
\end{align}
and the frequency-dependent decay rate \cite{shahbazyan-prb18}, 
\begin{equation}
\label{mode-decay-plas}
\gamma_{m}(\omega)=\dfrac{2\!\int \!  dV     
\varepsilon''(\omega,\bm{r})\bm{E}_{m}^{2}(\bm{r}) }{\!\int \!  dV     
[\partial\varepsilon'(\omega_{m},\bm{r})/\partial \omega_{m}]
 \bm{E}_{m}^{2}(\bm{r}) },
\end{equation}
where $\gamma_{m}(\omega)=-\gamma_{m}(-\omega)$. Note that   Eq.~(\ref{mode-coeff2}) is valid in the frequency region $\varepsilon''(\omega)/\varepsilon'(\omega)\ll 1$ or, equivalently, $\omega_{m}/\gamma_{m}\gg 1$.

The plasmon dyadic Green's function is given by $\bm{D}_{\rm pl}(\omega;\bm{r},\bm{r}')=\bm{\nabla}\bm{\nabla}' D_{\rm pl}(\omega;\bm{r},\bm{r}')$, where $D_{\rm pl}(\omega;\bm{r},\bm{r}')$ is defined by Eqs.~(\ref{green-exp}) and (\ref{mode-coeff2}),
\begin{equation}
\label{dyadic-plasmon0}
\bm{D}_{\rm pl}(\omega;\bm{r},\bm{r}') = \sum_{m}\frac{\omega_{m}^{2}}{2 U_{m}}\frac{\bm{E}_{m}(\bm{r}) \bm{E}_{m}  (\bm{r}')}{\omega_{m}^{2}-\omega^{2} -i\omega_{m}\gamma_{m}(\omega)}.
\end{equation}
Using Eqs.~(\ref{dissipation-coupling}) and (\ref{mode-decay-plas}), it is easy to check that the plasmon Green function (\ref{dyadic-plasmon0})  satisfies the relation 
\begin{align}
\label{optical}
\int \! dV  \varepsilon''(\omega,\bm{r}) \bm{D}_{\rm pl}^{*}(\omega;\bm{r},\bm{r}')
&
\bm{D}_{\rm pl}(\omega;\bm{r},\bm{r}'')
\nonumber\\
&
=4\pi \text{Im}\bm{D}_{\rm pl}(\omega;\bm{r}',\bm{r}''),
\end{align}
which is essential in the FD quantization approach.

For $\omega>0$, nonresonant contributions to $\bm{D}_{\rm pl}$ can be disregarded and the Green function takes the form
\begin{align}
\label{dyadic-plasmon1}
\bm{D}_{\rm pl}(\omega;\bm{r},\bm{r}') 
=
\sum_{m}
\frac{\omega_{m}}{4 U_{m}}\frac{\bm{E}_{m}(\bm{r}) \bm{E}_{m}  (\bm{r}')}{\omega_{m}-\omega -\frac{i}{2}\gamma_{m}(\omega)},
\end{align}
which satisfies the relation (\ref{optical}) as well. In the Markov approximation, i.e., $\gamma_{m}(\omega)\rightarrow \gamma_{m}(\omega_{m})\equiv \gamma_{m}$, the full Green functions (\ref{dyadic-plasmon0}) or (\ref{dyadic-plasmon1}) no longer satisfy the relation (\ref{optical}) but, near the resonance, the individual terms do. 

\section{Optical polarizability}

Consider a plasmonic system subjected to an incident monochromatic field $\bm{\mathcal{E}}_{i}e^{-i\omega t}$ that is uniform on the system scale. The near field generated by the plasmonic system in response to the incident has the form \cite{shahbazyan-prb18}
\begin{equation}
\label{mode-field-sc}
\bm{\mathcal{E}}(\omega,\bm{r})=\int \! dV'  \chi'(\omega,\bm{r}')\bm{D}_{\rm pl}(\omega;\bm{r},\bm{r}')\bm{\mathcal{E}}_{i}.
\end{equation}
Multiplying by Eq.~(\ref{mode-field-sc}) by $\chi'(\omega,\bm{r})$ and integrating over the system volume, we obtain the system induced dipole moment, $\bm{\mathcal{P}}=\int dV \chi'\bm{\mathcal{E}}$, as
\begin{equation}
\label{plasmon-induced-moment}
\bm{\mathcal{P}}(\omega)= \int \! dVdV'  \chi'(\omega,\bm{r})\chi'(\omega,\bm{r}')\bm{D}_{\rm pl}(\omega;\bm{r},\bm{r}')\!\cdot\! \bm{\mathcal{E}}_{i}.
\end{equation}
Inserting the plasmon Green function Eq.~(\ref{dyadic-plasmon0}) into Eq.~(\ref{plasmon-induced-moment}), we obtain
\begin{equation}
\bm{\mathcal{P}}(\omega)=\bm{\alpha}_{\rm pl}(\omega)\bm{\mathcal{E}}_{i}
\end{equation}
where $\bm{\alpha}_{\rm pl}(\omega)=\sum_{m}\bm{\alpha}_{m}(\omega)$ is the plasmon polarizability tensor \cite{shahbazyan-prb18}  and 
\begin{equation}
\label{polar-mode1}
\bm{\alpha}_{m} (\omega)=\frac{1}{\hbar}\frac{2\omega_{m}\,\bm{\mu}_{m}(\omega)\bm{\mu}_{m}(\omega)}{\omega_{m}^{2}-\omega^{2} -i\omega_{m}\gamma_{m}(\omega)},
\end{equation}
is the individual mode polarizability tensor while
\begin{equation}
\label{mode-matrix-element1}
\bm{\mu}_{m}(\omega)=\sqrt{\frac{\hbar\omega_{m}}{4U_{m}}}
\!\int\! dV\chi' (\omega,\bm{r})\bm{E}_{m}(\bm{r}) 
\end{equation}
is the plasmon optical transition matrix element. Near the resonance, the mode polarizability simplifies to 
\begin{equation}
\label{polar-mode2}
\bm{\alpha}_{m} (\omega)=\frac{1}{\hbar}\frac{\bm{\mu}_{m}(\omega)\bm{\mu}_{m}(\omega)}{\omega_{m}-\omega -\frac{i}{2}\gamma_{m}(\omega)}.
\end{equation}
Note that, in order to satisfy the optical theorem that guarantees energy flux conservation, the plasmon decay rate $\gamma_{m}(\omega)$ should also include the radiative decay contribution \cite{shahbazyan-prb18}. The latter is given by a standard expression for a point-like dipole
\begin{equation}
\gamma_{m}^{r}(\omega)=\frac{4\mu_{m}^{2}\omega^{3}}{3\hbar c^{3}},
\end{equation}
where $\omega$-dependence of $\mu_{m}$ is implied. In the Markov approximation, one should set $\omega=\omega_{m}$ in $\mu_{m}$ and $\gamma_{m}$.


\end{document}